\documentclass[letter,twocolumn]{jpsj3}
\usepackage{txfonts}
\usepackage{graphicx,color}
\usepackage{dcolumn}
\usepackage{bm}
\usepackage{tabularx}

\usepackage{physics}

\newcommand{\bpET}{$\beta^\prime$-(BEDT-TTF)$_{2}$ICl$_{2}$}
\newcommand{\TMX}{(TMTTF)$_2$\textit{X}}

\title{Stability of Correlated Insulating States in Molecular Conductors from First-Principles Calculation}

\author{Takao Tsumuraya$^{1,2}$\thanks{E-mail:tsumu@kumamoto-u.ac.jp}, Tsuyoshi Miyazaki$^3$, and Hitoshi Seo$^{4,5}$\thanks{E-mail:seo@riken.jp}}
\inst{$^1$Magnesium Research Center, Kumamoto University, Kumamoto 860-8555, Japan\\
$^2$Department of Materials Science and Engineering, Kumamoto University, Kumamoto 860-8555, Japan\\
$^3$Research Center for Materials Nanoarchitectonics~(WPI-MANA),
National Institute for Materials Science, Tsukuba, Ibaraki 305-0044, Japan\\
$^4$Condensed Matter Theory Laboratory, RIKEN, Wako-shi, Saitama 351-0198, Japan\\
$^5$RIKEN Center of Emergent Matter Science (CEMS), Wako-shi, Saitama 351-0198, Japan}

\abst{Electronic properties of molecular conductors exhibiting antiferromagnetic (AFM) spin order and charge order (CO) owing to electron correlation are studied using first-principles density functional theory calculations. 
We investigate two systems, a quasi-two-dimensional Mott insulator \bpET~with an AFM ground state, and several members of quasi-one-dimensional \TMX~showing CO. 
The stabilities of the AFM and CO states are compared between the use of a standard exchange-correlation functional 
based on the generalized gradient approximation and that of a range-separated hybrid functional; 
we find that the latter describes these states better. 
For \bpET, the AFM order is much stabilized with a wider band gap. 
For \TMX, only by using the hybrid functional, the AFM insulating state is realized and the CO states coexisting with AFM order are stable under structural optimization,
 whose stability among different \textit{X} shows the tendency consistent with experiments. }

\begin{document}
\maketitle

Molecular conductors are studied as typical correlated electron systems, taking advantage of their simple electronic structure near the Fermi level despite their complicated crystal structures~\cite{Lebed_book}. 
They present fundamental many-body problems, such as the interplay between strong electron-electron interaction and the underlying lattice structure controlled by molecular arrangements, together with relatively strong electron-phonon interactions, which can lead to symmetry breaking. 
A representative issue is the competition between antiferromagnetic (AFM) ordering and quantum-disordered states such as the spin-gapped or spin-liquid states~\cite{Kanoda_Kato_2011ARCMP}. 
Another is the role of the lattice degree of freedom in stabilizing charge ordered (CO) states whose primary driving force is considered to be the electron-electron repulsion~\cite{Seo_JPSJ2006}. 

To elucidate the mechanisms of their rich physical properties, effective low-energy models such as the Hubbard-type models have extensively been applied~\cite{Seo_ChemRev,Powell_2011}. 
Their microscopic parameters can now be evaluated in a quantitative level using first-principles band calculations based on the density-functional theory (DFT), thanks to the modern computational developments making accessible to large number of atoms in the unit cell~\cite{Miyazaki_1995,Xu_1995,Miyazaki_PRB96_DCNQI,k_ET_KNakamura_JPSJ09,Yoshimi_PRL2023,Ishibashi_STAM_review09}. 
On the other hand, standard first-principles calculations often run into essential problems when attempting to directly reproduce insulating states of molecular conductors due to symmetry breaking, 
 such as the aforementioned AFM and CO states.

This is related to the fact that standard treatments using local density approximation~\cite{BarthHedin_1972} 
or generalized gradient approximation (GGA)~\cite{GGA_PBE} have a problem of unphysical self-interaction error~\cite{PerdewZunger1981}
and generally underestimate the band gaps.
To overcome this difficulty, in studies of inorganic strongly correlated systems such as transition metal oxides, 
the ``+$U$" method has been heavily used~\cite{Anisimov_1991}, implementing the strong correlation effect in the $d$/$f$ atomic orbitals. 
However, this method cannot be simply adopted to molecular conductors whose electronic properties are governed by molecular orbitals, 
on which ``on-site'' Coulomb repulsion acts~\cite{KinoFukuyama_1996}.
Here we address this issue by choosing an alternative approach, rather than implementing the strong Coulomb repulsion effect on the molecular orbitals,
but to improve the treatment of the exchange-correlation term in DFT. 

In this context, 
 we have shown the efficiency of the use of a range-separated hybrid functional 
 by Heyd, Scuseria, and Ernzerhof (HSE)~\cite{HSE03,HSE_Martin_JCP}, 
 for several molecular CO systems~\cite{D3Cat_tsumu,Tsumu_cryst21}.
In the hybrid functional method, the exchange-correlation energy is evaluated by mixing the exact Fock exchange and exchange-energy functional used in LDA or GGA, 
 whose range-separated formalism was proposed by HSE for the description of solid states. 
In Refs.~\citen{D3Cat_tsumu} and~\citen{Tsumu_cryst21}, 
the structural stability of the CO state
was demonstrated, in clear contrast with the standard approaches in which CO structures are unstable and reduced to the non-CO state by the structural optimization process. 

\begin{figure}
    \centering
    \includegraphics[width=0.95\linewidth]{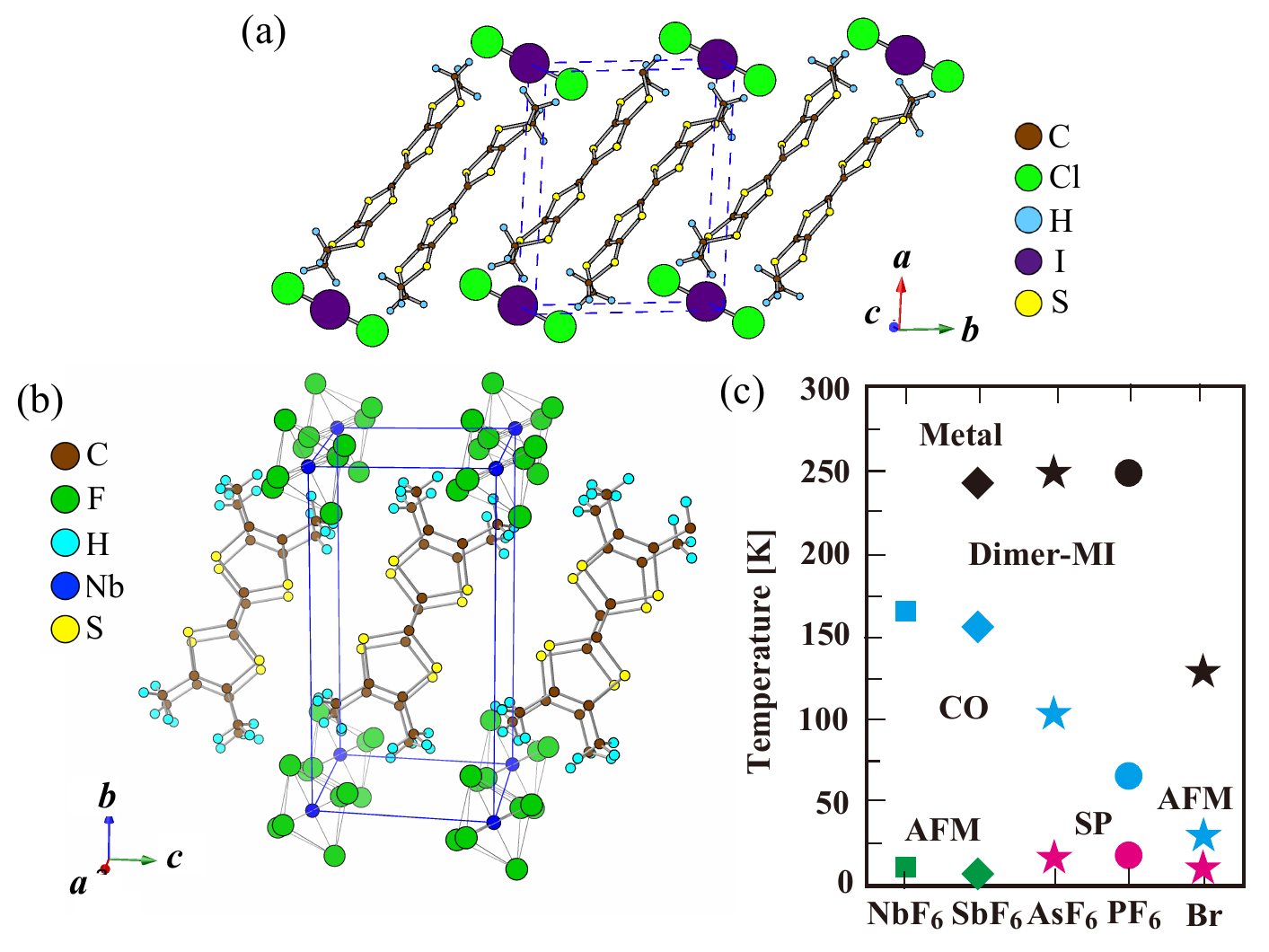}
    \caption{(Color online)~Crystal structures of (a) \bpET \ and (b) \TMX. (c) A schematic experimental phase diagram of the \TMX \ compounds:  
    The horizontal axis corresponds to the chemical pressure with smaller anion to the right. 
    The symbols represent the crossover temperature to the dimer-Mott insulating (dimer-MI) phase~\cite{note_NbF6}, 
    the transition temperatures to charge ordered (CO), antiferromagnetic (AFM), and spin-Peierls (SP) phases, respectively.} 
    \label{fig1}
\end{figure}

In this paper, we investigate the correlation-induced insulating states further taking account of AFM ordering.
AFM insulating states are ubiquitously seen in molecular conductors and are extensively studied within effective model approaches~\cite{KinoFukuyama_1996,Seo_Fukuyama_1D_1997,Seo_2000,Kino_Beta_ET2Cl2,Naka_2019,Naka_2020, CLAY2019}; 
however, due to the difficulties mentioned above, they have not been properly studied by first-principles calculations even for typical compounds. 
In the following, we first investigate the insulating state with AFM order in \bpET~\cite{Kobayashi_betap_ET2ICl2,Taniguchi_Betap_ICl2}, 
 which is a Mott insulator with an AFM transition temperature $T_\textrm{N}=22$~K~\cite{Tokumoto_1987, Yoneyama1997}.  
Then we discuss a series of \TMX, showing CO 
 whose transition temperature ($T_{\rm CO}$) varies by changing the anion as 
\textit{X} = NbF$_6$ ($T_{\rm CO}=165$~K), AsF$_6$ (100~K), 
 and PF$_6$ (70~K)~\cite{Chow_2000,Monceau_2001,Zamborszky_2002,Kitou_2017, Kitou_2021}. 
In (TMTTF)$_2$NbF$_6$, at the ground state, CO and AFM spin order ($T_\textrm{N}=10$~K) coexist, while the other two compounds undergo a spin-Peierls transition to spin-gapped states accompanying lattice distortions.
The crystal structures of these two systems are shown in Figs.~\ref{fig1}(a) and \ref{fig1}(b), respectively, 
 and a schematic phase diagram for the \TMX \ including the salts studied in this paper is drawn in Fig.~\ref{fig1}(c).
Here we investigate the systems without considering the lattice distortions due to the spin-Peierls transition and investigate the AFM state as its non-distorted counterpart.

We perform first-principles calculations based on plane-wave basis sets and the pseudopotential technique with the projector-augmented wave method~\cite{Vanderbilt1990, PAW1994}, 
 within the Vienna $ab$ $initio$ simulation package~\cite{Kresse_VASP1996,Kresse_Joubert_PAW1999}. 
Spin-dependent exchange-correlation functionals are used, and 
we compare the standard GGA functional by Perdew, Burke, and Ernzerhof (PBE)~\cite{GGA_PBE}
 and the hybrid functional by HSE (HSE06)~\cite{HSE_Martin_JCP}. 
The cutoff energies for plane waves are set to 700 eV for both the PBE and HSE06 functionals.
The ${\bm k}$-point meshes for $\beta^\prime$-(BEDT-TTF)$_2$ICl$_2$ were set to 4$\times$6$\times$2 and 4$\times$3$\times$2, and for \TMX, 4$\times$4$\times$2 and 4$\times$2$\times$2, for the nonmagnetic (NM) and AFM states, respectively.
The density of states (DOS) for the AFM states are calculated using the ${\bm k}$-point meshes of 4$\times$4$\times$2.

As for the experimental structures, we used that of  
\bpET \ measured at 12~K~\cite{MWatanabePV, MatsunagaMThesis04}, 
and those of \TMX \ at 30~K~\cite{Kitou_2021}.
The positions of the H and F atoms are optimized by the first-principles calculations.
The data for \TMX \ are within the CO phase, with the so-called ferroelectric CO pattern, 
 losing the inversion symmetry that the system holds above $T_{\rm CO}$~\cite{Monceau_2001,Yoshimi_TM_PRL2012}, 
 but above the spin-Peierls transition temperatures for $X$ = AsF$_6$ and PF$_6$. 
We also performed structural optimization for the internal coordinates, 
fixing the lattice parameters and relaxing all the atomic positions. 

Experimentally, $\beta^\prime$-(BEDT-TTF)$_2$ICl$_2$ exhibits insulating behavior below room temperature.~\cite{Kobayashi_betap_ET2ICl2}. 
NM band calculations show that their conduction band is half-filled~\cite{Miyazaki_PRB_betap_ET2ICl2, Koretsune_Hotta_PRB2014},  
 therefore the compound is considered as a Mott insulator.  
Since the nominal charge of BEDT-TTF is $+1/2$ and the dimerized structure (see Fig.~\ref{fig1}) brings about the half-filling 
 of the antibonding combination of the HOMO between BEDT-TTF forming dimers, 
 it is a typical dimer-Mott insulator. 
A theoretical analysis based on the Hubbard model describing the HOMO level of BEDT-TTF with the $\beta^\prime$-type arrangement shows the stability of the AFM order~\cite{Kontani_2003,Kino_Beta_ET2Cl2}. 

\begin{figure}
\begin{center}
\includegraphics[width=0.95\linewidth]{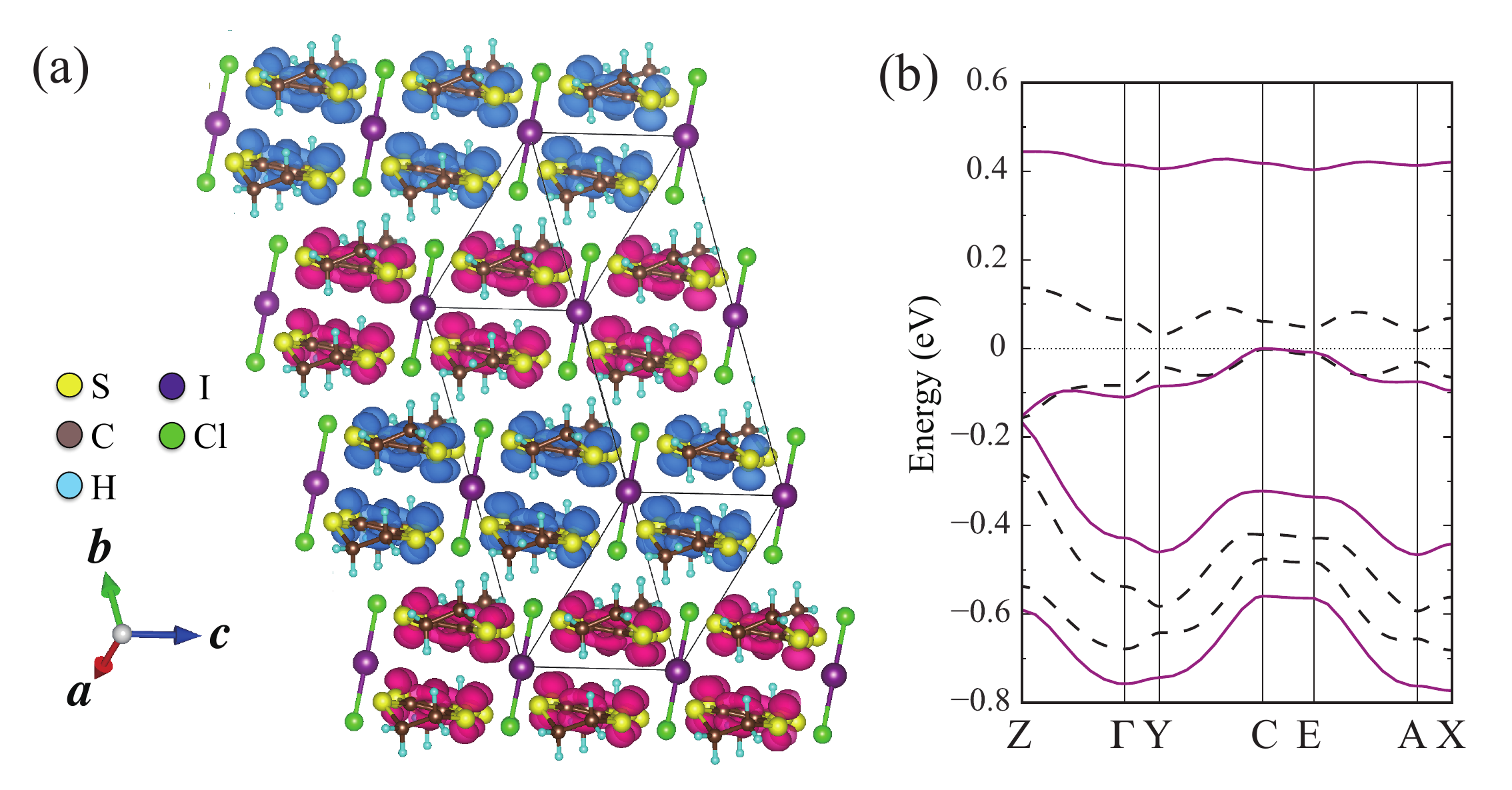}
\end{center}
\caption{(Color online)~(a)~Spin density distribution of the antiferromagnetic state of $\beta^\prime$-(BEDT-TTF)$_2$ICl$_2$, 
calculated using the HSE06 functional. 
(b) Its band structures, where the dashed and solid curves are the results with the PBE and HSE06 functionals, respectively. The dotted horizontal lines at 0~eV show the top of the valence bands. }
\setlength\abovecaptionskip{0pt}
\label{bands_ET2ICl2_AFM}
\end{figure}

Using the experimental structure, we find an AFM solution with a unit cell twice of the primitive cell along the $b$ axis which contains four BEDT-TTF molecules. 
The spin-dependent self-consistent field calculations show an AFM ordering between the dimers. 
Within the PBE functional, the magnetic moments are evaluated as 0.31 $\mu_{\rm B}$/dimer, while using the HSE06 functional, 
 they become 0.58 $\mu_{\rm B}$/dimer, as listed in Table~\ref{Moment_TM};  
these are calculated by the sum of magnetic moments on the C and S atoms. 
The spin density distribution is shown in Fig.~\ref{bands_ET2ICl2_AFM}(a) 
 using the results for the HSE functional, whose spin pattern is common with those within PBE:
the dimers with spin-up and spin-down electrons alternate along the $b$-axis. 
This pattern is consistent with the theoretical results using the Hubbard model~\cite{Kontani_2003,Kino_Beta_ET2Cl2}. 
The dashed curves in Fig.~\ref{bands_ET2ICl2_AFM}(b) show the band structure of the AFM state within PBE, which exhibits a band gap of 0.06 eV.
Notably, the band gap expands to 0.42 eV when 
we use the HSE06 functional, as shown by the solid curve in Fig.~\ref{bands_ET2ICl2_AFM}(b).
We also perform structure optimizations using the HSE functional with the AFM ordering, 
 which show the band gap and the magnetic moments of 0.39 eV and 0.57 $\mu_{\rm B}$/dimer, respectively, close to the results above.  

\begin{table}
\begin{center}
\caption{Magnetic moments per molecular dimer (in $\mu_B$) calculated by using the PBE and HSE06 functionals, for the experimental structures. 
For the TMTTF salts, the values for the two monomers in the dimer are also listed.}
\label{Moment_TM}
\vspace{1mm}
\small
\begin{tabular}{l|c|cc|c|cc}
\hline\hline
 Functional & \multicolumn{3}{c|}{GGA-PBE} & \multicolumn{3}{c}{HSE06} \\
 \hline
 & dimer  & \multicolumn{2}{c|}{monomer} & dimer  & \multicolumn{2}{c}{monomer} \\
  \hline
$\beta^\prime$-ET$_2$ICl$_2$ 
& 0.31 & & & 0.58 &\\
\hline
(TMTTF)$_2$NbF$_6$ & 0.06 & 0.05 & 0.01 & 0.48 & 0.33 & 0.15\\
(TMTTF)$_2$AsF$_6$ & -- & -- & -- & 0.51 & 0.30 & 0.21\\
(TMTTF)$_2$PF$_6$ & -- & -- & -- & 0.49 & 0.25 & 0.24\\
\hline
\hline
\end{tabular}
\end{center}
\end{table}

Experiments on TMTTF salts reveal insulating ground states, despite the calculated NM band structure showing a metallic quarter-filled band at the Fermi level.~\cite{Jacko_TMTTF_2013,Seo_ChemRev}. 
Therefore, the correlation effect is considered to be crucial. 
The weakly dimerized crystal structure brings about an effectively half-filled band and the tendency toward the dimer-Mott insulating state owing to the electron correlation, i.e., 
 the on-site (intra-molecular) Coulomb repulsion in terms of the quarter-filled Hubbard model with dimerization~\cite{Emery_1982}. 
In addition, the intersite (inter-molecular) Coulomb repulsion terms are considered to be responsible for the CO instability~\cite{Seo_Fukuyama_1D_1997}, 
while the role of electron-phonon interaction has been investigated~\cite{CLAY2019,Yoshioka_2012}. 
We note that the electronic properties of \textit{X} = NbF$_6$ 
 are similar to those for \textit{X} = SbF$_6$~\cite{Yu2004_SbF6}, with $T_{\rm CO} = 157$~K and $T_{\rm N}=6$~K, 
 on which there are more experimental studies conducted;  
the calculations for \textit{X} = SbF$_6$ were not performed here due to the lack of structural data for the low temperature CO phase. 

\begin{figure}
\begin{center}
\includegraphics[width=0.95\linewidth]{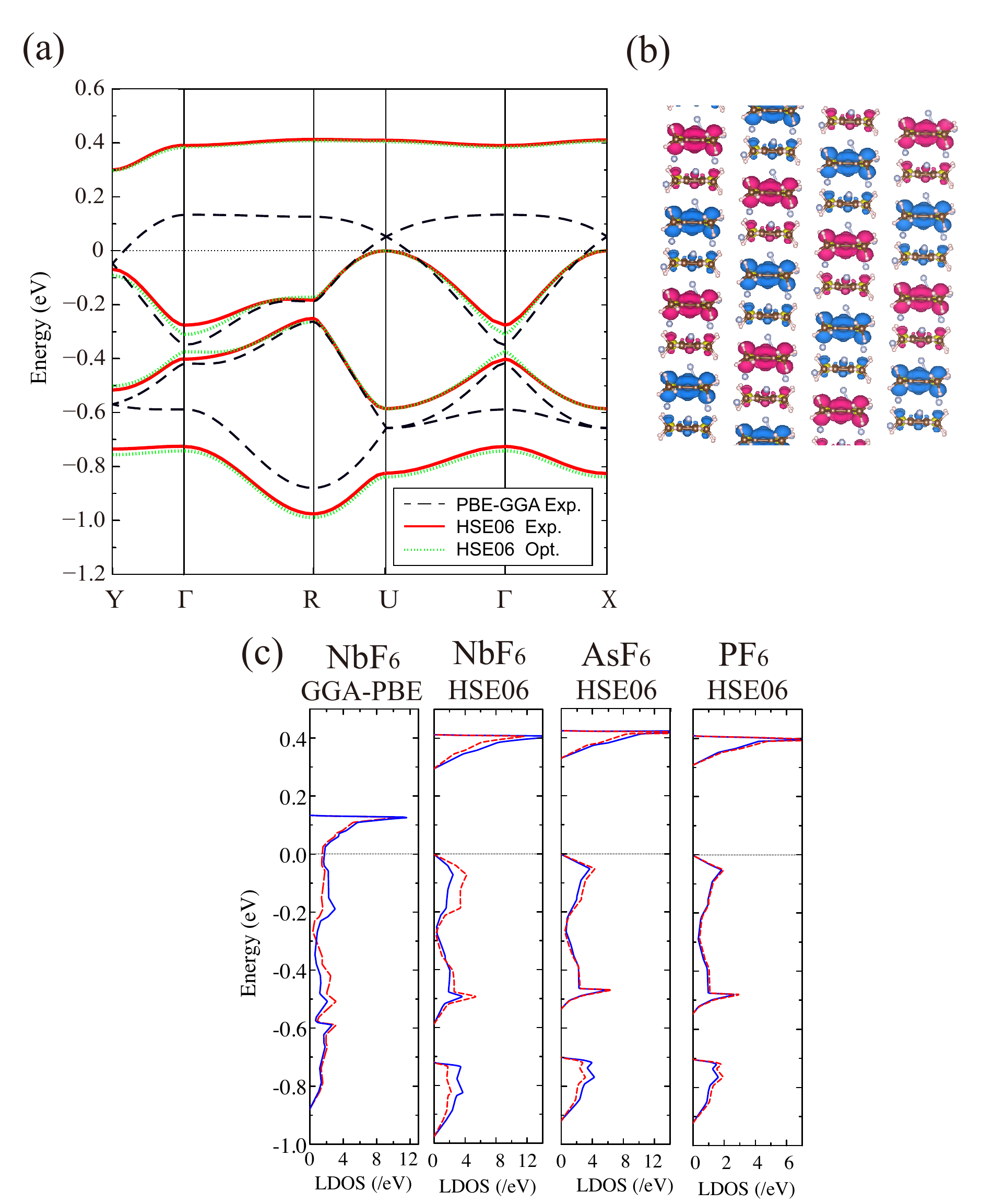}
\end{center}
\setlength\abovecaptionskip{-2pt}
\caption{(Color online)~(a) Band structures for AFM + CO phase in (TMTTF)$_2$NbF$_6$, calculated by GGA-PBE and HSE06 functionals.  
The origin in the longitudinal axis (dotted lines) 
is taken as the Fermi energy and the top of the valence bands, respectively. The green dotted curves depict the HSE06 band structure for the experimental structure. 
(b) Calculated spin densities with the HSE06 functional. (c) Local DOS for \textit{X} = NbF$_6$, AsF$_6$, and PF$_6$, in which geometric optimization is performed with the HSE06 functional.
The LDOS calculated with GGA-PBE for NbF$_6$ salt is based on the experimental structure.}
\label{disp_30K_NbF6}
\end{figure}

Here we seek for AFM self-consistent solutions with a unit cell twice of the primitive cell which contains four TMTTF molecules, 
 suggested in the literature~\cite{Yoshimi_TM_PRL2012,Yoshimi_PRL2023}. 
For the experimental structure, 
 using PBE, the AFM state was achieved for \textit{X} = NbF$_6$, whereas for \textit{X} = AsF$_6$ and PF$_6$, 
 only NM states were obtained within the numerical accuracy. 
The results for \textit{X} = NbF$_6$ indicate the magnetic moments on the TMTTF molecules in each dimer to be 0.01~$\mu_{\rm B}$ and 0.05~$\mu_{\rm B}$ (Table~\ref{Moment_TM}). 
The band structure is depicted in Fig.~\ref{disp_30K_NbF6}(a), showing a metallic state, 
but with negligible charge disproportionation between the two molecules as seen from Fig.~\ref{disp_30K_NbF6}(c). 

When the hybrid functional is applied, we can now obtain AFM solutions in all the three salts, 
 coexisting with CO, showing different degree of spin polarization in the two TMTTF molecules within each dimer.  
Moreover, they all show insulating states; the band structure for \textit{X} = NbF$_6$ 
 is depicted in Fig.~\ref{disp_30K_NbF6}(a) showing a clear band gap of 0.30~eV. 
We obtain insulating states for \textit{X} = AsF$_6$ and PF$_6$ as well with band gaps of 0.33~eV and 0.34~eV, respectively, 
 in contrast to the case of PBE where only metallic states are found for all the compounds. 
The AFM pattern coexisting with CO is drawn in Fig.~\ref{disp_30K_NbF6}(b), 
 and the estimated magnetic moments are listed in Table~\ref{Moment_TM}, together with the PBE results. 
While the values per dimer are similar for the three salts (about 0.5~$\mu_{\rm B}$), 
 their difference between the two TMTTF molecules is largest in \textit{X} = NbF$_6$ (0.18~$\mu_{\rm B}$), 
 next in \textit{X} = AsF$_6$ (0.09~$\mu_{\rm B}$), and almost vanishes in \textit{X} = PF$_6$ (0.01~$\mu_{\rm B}$). 
This is compatible to the degree of CO and with the transition temperatures $T_{\rm CO}$, 
 whose structural stability will be investigated next. 

As mentioned in the introduction, when we use PBE functional, 
 the CO structure is lost after the structural optimization process. 
Namely, the structural/electronic difference between the two TMTTF molecules turns negligibly small; 
therefore in the following we show results for the HSE06 functional. 
First, for \textit{X} = NbF$_6$, the CO state is properly kept. 
From the structural viewpoint, we can see this from the difference in the lengths of the central C=C bond in the two molecules, as listed in Table~\ref{ccbonds_TM}. 
Although the calculated difference, 0.007~\AA, \ is smaller than the experimental value, 0.034~\AA, 
the results show a stable CO structure. 
This difference becomes smaller for \textit{X} = AsF$_6$ (0.002~\AA) and numerically negligible in \textit{X} = PF$_6$. 
Considering the tendency that the structural stability is underestimated in other CO systems as well~\cite{Tsumu_cryst21,D3Cat_tsumu}, 
 the variation among different \textit{X} is consistent with the experimental CO stability, 
 e.g. reflected by the transition temperatures $T_{\rm CO}$ (see Fig.~\ref{fig1}). 

\begin{table}
\begin{center}
\caption{Anion dependence of the distances of the central C=C bond in (TMTTF)$_2$\textit{X}, 
comparing the results after structural optimization using HSE06 functional and experimental data at 30~K~\cite{Kitou_2021}.}
\label{ccbonds_TM}
\begin{tabular}{cc|ccc}
\hline\hline
\  &  HSE06 & Exp. \\
 \   &  $d$(C=C)[\AA] & $d$(C=C)[\AA]\\
 \hline
(TMTTF)$_2$NbF$_6$ & 1.371 & 1.389 \\
        & 1.364 & 1.355 \\
(TMTTF)$_2$AsF$_6$ & 1.369 & 1.396 \\
        & 1.367 & 1.352 \\
(TMTTF)$_2$PF$_6$ & 1.368 & 1.392 \\
       & 1.368 & 1.371 \\
\hline
\hline
\end{tabular}
\end{center}
\end{table}

To see the degree of charge disproportion in the optimized structure, 
 we plot the local density of states (LDOS) in Fig.~\ref{disp_30K_NbF6}(c), 
 which is obtained as a summation of the projected DOS on the orbitals of C and S atoms. 
The red (blue) lines correspond to LDOS for the TMTTF molecule with shorter (longer) central C=C bond length; 
 the results for the three salts are shown together with the results using PBE for \textit{X} = NbF$_6$ 
 within the experimental structure 
 for comparison. 
In the occupied states of \textit{X} = NbF$_6$ with the energy range from --0.6 to 0~eV, 
 there is a clear difference between the LDOS. This indicates the molecules with the shorter bond are charge-rich. 
On the contrary, from --1.0 to --0.7~eV and from 0.3 to 0.4 eV, the LDOS of the charge-poor molecules is slightly larger. 
These results clearly show the charge disproportionation between the molecules. 
Such a difference becomes obscure for \textit{X} = AsF$_6$ and almost absent in \textit{X} = PF$_6$;  
this reduction corresponds to the systematic variation in the bond lengths presented in Table~\ref{ccbonds_TM}. 

Here we compare our results with the experimental results. 
In $\beta^\prime$-(BEDT-TTF)$_2$ICl$_2$, from the temperature dependence of the resistivity, 
 the activation energy is evaluated as 0.113~eV~\cite{Taniguchi_Betap_ICl2}, 
 consistent with an optical measurement~\cite{Hashimoto_betap_ET2ICl2}. 
Our results show a direct band gap of 0.06~eV using PBE functional, and 0.4 eV using HSE06; 
 although the comparison with the Mott gap experimentally opened above $T_{\rm N}$ is difficult, 
 we see that PBE (HSE06) underestimates (overestimates) the gap. 
As for the magnetic moments, NMR suggests the ordered moment at almost 1~$\mu_{\rm B}$ per dimer at low temperatures~\cite{Eto_betap_ET2ICl2}. 
The calculated results of 0.31~$\mu_{\rm B}$ for PBE and 0.58~$\mu_{\rm B}$ for HSE06 suggest that the latter provides closer estimates. 

Systematic evaluation of activation energy in TMTTF salts by conductivity measurements gives 
0.043,  0.031, and 0.032~eV, for \textit{X} = SbF$_6$, AsF$_6$, and PF$_6$, respectively~\cite{Nad_2006}. 
Here, instead of \textit{X} = NbF$_6$ whose data were not available, we refer to the value for \textit{X} = SbF$_6$ that shows similar 
 properties and transition temperatures as mentioned above. 
In our calculations insulating states could only be obtained by the HSE06 functional, whose band gaps are 0.30, 0.33, and 0.34~eV for \textit{X} = NbF$_6$, AsF$_6$, and PF$_6$, respectively. 
Again, although the comparison with the gap seen in experiments, opened well above $T_{\rm N}$, 
 may not be appropriate, the same tendency that PBE (HSE06) underestimates (overestimates) the gap is seen. 
We should also note that in \TMX, the quasi-one-dimensional electronic structure is expected to bring about low-dimensional physics when electron correlation plays a role, 
 which we need to be careful about when making the comparison between our first-principles evaluation and experiments. 
The low dimensionality is known to affect the magnetic ground state as well, 
 evidenced from the spin-Peierls instability in \textit{X} = AsF$_6$ and PF$_6$ accompanying lattice distortion~\cite{Kuwabara_2003,Otsuka_2008,Yoshioka_2012, Clay_PRB2003, Clay_PRB2007, CLAY2019}, 
 not considered in our study. 
Although we could find stable AFM solutions coexisting with CO in all the three salts within HSE06, 
 the comparison of the magnetic state is available only for \textit{X} = NbF$_6$. 
Again, due to the absence of data for \textit{X} = NbF$_6$, we refer to an NMR estimate for \textit{X} = SbF$_6$~\cite{Nomura_2015,Matsunaga_2013}: 
 0.70~$\mu_{\rm B}$ and 0.24~$\mu_{\rm B}$ for the charge rich and poor TMTTF molecules, respectively. 
As shown in Table~\ref{Moment_TM}, the calculated values are smaller: 0.33~$\mu_{\rm B}$ and 0.15~$\mu_{\rm B}$.

In summary, we investigate the stability of insulating states caused by symmetry breakings in correlated molecular conductors
by using the exchange-correlation functional based on the standard GGA (PBE) 
 and a range-separated hybrid functional (HSE06). 
The antiferromagnetic insulating state of \bpET~is reproduced with larger band gap and magnetic moments with HSE06. 
In (TMTTF)$_2$\textit{X}, we fail to obtain the charge ordered state by PBE, 
but succeeded by the use of HSE06; we show a stable charge ordered electronic state coexisting with antiferromagnetic order, 
 seen experimentally in \textit{X} = NbF$_6$. 
The structural relaxation in the TMTTF salts using HSE06, although the degree of charge order is depressed, 
 shows the stability of charge ordered state seen structurally and electronically with distinct TMTTF molecules in the unit cell.
We note that such structural stability is important in a sense that now we have a solid ground to calculate the dynamical motions of atoms, 
such as the molecular vibrations which have been providing rich information in many experimental studies. 

\bibliographystyle{jpsj}
\bibliography{./library}

\section{ACKNOWLEDGEMENTS}
JSPS Grant-in-Aid funded this research for Scientific Research Nos.~16K17756, 20H04463, 20H05883, 23H01129, 23K03333, and 23H04047. 
Computational work was performed under the Inter-university Cooperative Research Program and the Supercomputing Consortium for Computational Materials Science of the Institute for Materials Research (IMR), Tohoku University (Proposals No.~202212-RDKGE-0048). 
The calculations were mainly performed using the MASAMUNE supercomputer system at IMR, Tohoku University, and the ``Flow" supercomputer at Information Technology Center, Nagoya University. 
TM thanks the support by World Premier International Research Center Initiative (WPI), MEXT, Japan.

\end{document}